# Brain-scale Theta Band Functional Connectivity As A Signature of Slow Breathing and Breath-hold Phases

Anusha A. S., Pradeep Kumar G., and A. G. Ramakrishnan

*Abstract*—The study reported herein attempts to understand the neural mechanisms engaged in the conscious control of breathing and breath-hold. The variations in the electroencephalogram (EEG) based functional connectivity (FC) of the human brain during consciously controlled breathing at 2 cycles per minute (cpm), and breath-hold have been investigated and reported here. An experimental protocol involving controlled breathing and breath-hold sessions, synchronized to a visual metronome, was designed and administered to 20 healthy subjects (9 females and 11 males, age: 32.0 ± 9.5 years (mean ± SD)). EEG data were collected during these sessions using the 61-channel eego$^{TM}$mylab system from ANT Neuro. Further, FC was estimated for all possible pairs of EEG time series data, for 7 EEG bands. Feature selection using a genetic algorithm (GA) was performed to identify a subset of functional connections that would best distinguish the inhale, exhale, inhale-hold, and exhale-hold phases using a random committee classifier. The best accuracy of 93.36 % was obtained when 1161 theta-band functional connections were fed as input to the classifier, highlighting the efficacy of the theta-band functional connectome in distinguishing these phases of the respiratory cycle. This functional network was further characterized using graph measures, and observations illustrated a statistically significant difference in the efficiency of information exchange through the network during different respiratory phases.

## I. INTRODUCTION

Breathing is a fundamental process of the human body and is central to life. Although it is mostly an involuntary and rhythmic act that continues during sleep and even when a person is unconscious, it is voluntary when a person controls breathing during talking, singing, or even during voluntary breath-holding. Several years of research have led to the understanding that these involuntary and voluntary control systems emerge from separate sites in the central nervous system and have separate descending pathways [1]. Involuntary control is mediated by both rhythmic and nonrhythmic neuronal assemblies in the hindbrain (the pons and medulla breathing control centres in the brain stem). Voluntary control of breathing, on the other hand, emerges from the motor and premotor cortex, stimulating the neurons that innervate the diaphragm and intercostal muscles and is mediated by the descending corticospinal tract. Thus, although the process of breathing with its rhythmicity and seemingly trivial nature urges one to think that the underlying neural mechanisms are straightforward, they are not so actually.

While most of the existing literature on the neurophysiology of breathing has focused on autonomic brain stem processes [1]–[3], the higher brain mechanisms underlying the cognitive aspects of breathing have not been explored much. Recent years have seen an increasing interest in understanding breathing as a complex behaviour, involving cortical areas above the brain stem, with surprising facets that go beyond facilitating gas exchange [4]. Evidence of breathing-related entrainment occurring in cortical neural networks that are critical to emotion, cognition, and memory from animal studies [5], [6] also serves as a motivation to explore further on similar lines.

This study attempts to understand functional connectivity variations in the human brain during breathing and breath-hold using electroencephalography (EEG). The inhalation and exhalation phases of the breathing cycle as well as the inhale-hold and exhale-hold phases during voluntary breath-holding are analysed to see if the brain re-wires itself during these phases to assign distinct functional connectivity (FC) signatures for them. Such signatures are a major focus in the emerging fields of brain-body neuroscience and interoception. This newly emerging domain brings with it unique opportunities for computational, clinical, and translational neuroscience investigations probing the breadth, nature, and extent of breath-brain interactions. This study is envisaged as a precursor to such investigations and it attempts to explore how breathing rhythms modulate brain-scale functional connections.

## II. PREVIOUS WORK

Breathing is indeed a "breathtaking" process that orchestrates the whole symphony of life in all living organisms. Fundamentally, it supports life by facilitating the exchange of oxygen and carbon dioxide into and out of lungs, relentlessly and with incredible efficiency, to as much as 672,768,000 breaths over a human life span of 80 years. The rather automatic

Anusha A. S. and Pradeep Kumar G. are with the Department of Electrical Engineering, Indian Institute of Science, Bengaluru, India, e-mail: anushas@iisc.ac.in, pradeepkg@iisc.ac.in.

A. G. Ramakrishnan is with the Department of Electrical Engineering, Indian Institute of Science, Bengaluru, India, Centre for Neuroscience, Indian Institute of Science, Bengaluru, India, and Heritage Science and Technology, Indian Instiute of Technology Hyderabad, Hyderabad, India e-mail: agr@iisc.ac.in.

and subconsciously controlled nature of the breathing process might prompt one to think that the underlying mechanisms are also elementary and uncomplicated. However, a wealth of literature concerning the physiological and neurophysiological mechanisms of breathing proves the contrary. Researchers, over the years, have started to realize that breathing involves an exceptionally robust and complex mechanism with surprising facets that go beyond just enabling the gas exchange. A summary of the recent and relevant literature on breath research is presented here.

*A. Studies on Spontaneous Breathing, Paced Breathing, and Breath-hold*

As already mentioned, spontaneous breathing in mammals is steered and regulated by neural assemblies in the brainstem, extending from the pons to the lower medulla. In humans, these serially assembled neural circuits in the brainstem functionally integrate at multiple hierarchical levels to generate and maintain flexible, state-dependent respiratory rhythms, thereby enabling a wide repertoire of respiratory patterns [7]. Recent studies have also highlighted the significant role of some suprapontine structures such as the diencephalon, the limbic areas, the striatum, and the cortex in the modulation of respiratory drive of the brainstem [8]. The involuntary breathing pattern generated and regulated by the respiratory centers in the pons and medulla ensures that the oxygen, carbon dioxide, and pH levels of blood are maintained within acceptable limits. However, human beings are capable of volitional breath control during various activities like speaking, singing, deep diving, or crying. Evidence suggests that such active control of breathing is influenced by stimulation of more complex neural circuitry above the brain stem, involving the hippocampus [5], amygdala [9], and insula [10].

Effects of voluntary breath-hold on whole-brain rhythms were explored by some researchers using different modalities like EEG, magnetic resonance encephalography (MREG), intracranial EEG (iEEG), and functional magnetic resonance imaging (fMRI) [11]–[16]. These studies were based on the hypothesis that breath-hold related elevation in cerebral blood flow alters cardiovascular brain pulsations due to changes in vasomotor tone in response to breath-hold.

Raitamaa *et al.* observed a dynamic increase in MREG cardiovascular brain pulsation amplitudes in cortical areas and sagittal sinus of the brain during repeated breath-holds [11]. The study highlighted the efficacy of MREG scanning as a tool to evaluate brain activity with respect to changes in physiological brain pulsations. They also reported that the spatial propagation of averaged cardiovascular impulses varied as a function of successive breath-hold trials, exhibiting a decreasing spatial similarity over time. Effects of breath-hold on the EEG global field power (GFP) and regional field power (RFP) of different cortical areas above the brain stem were analysed and increased brain activity was observed in the middle central and middle anterior regions, with the wave of cortical activation further spreading out to more lateral and posterior areas of the brain [12]. Furthermore, the linear and nonlinear coupling between EEG signal (delta and alpha bands) and variations in end-tidal carbon dioxide ($P_{ET}CO_2$) levels during breath-hold were evaluated using two indices, namely maximal information coefficient (MIC, a generic measure of linear and nonlinear correlations) and MIC-$\rho^2$ (a "pure" nonlinear correlation measure) [13], [14]. While these indices in the delta band could discriminate free-breathing from breath-hold in several cortical regions, in the alpha band, the same pattern was observed only for the MIC-$\rho^2$ index. These results were therefore an indication of a generic coupling between delta EEG power and $P_{ET}CO_2$ and a pure nonlinear interaction between alpha EEG power and $P_{ET}CO_2$. Moreover, higher values of indices were found for the breath hold task with respect to free breathing. A similar study evaluated both cortical and subcortical brain activity related to hypercapnia using fMRI [15] and the results highlighted a supra-linear response to $CO_2$ challenges in the brainstem, thalamus, and putamen.

Previous studies from our group [16], [17] had analysed the effects of slow breathing and breath-hold on whole-brain functional connectivity and illustrated that the brain FC exhibits a hemispherical symmetry during breath-hold in the delta and alpha bands [16]. Synchronization of neuronal assemblies in different cortical regions of the brain was found to be higher in low-frequency EEG bands and lower in high-frequency EEG bands.

*B. Studies on Breathing As An Emotion Barometer*

Even though breathing is primarily regulated for physiological homeostasis and metabolism, it can also change in response to emotions (like happiness, sadness, fear, anxiety) and cognitive states (like alertness, stress, attention) [18]. Accordingly, understanding the effects of breathing on emotion and cognition as well as its therapeutic potential has been an important focus area in breath research. Preliminary evidence from animal studies also suggests a profound impact of breathing on emotional and cognitive states [5], [6].

Yackle *et al.* identified a neuronal subpopulation in the mouse respiratory centre that can strike a balance between calm and arousal behaviors throughout the brain [19]. The study demonstrated that slow, controlled breathing can induce tranquility, and can even suppress excessive arousal patterns such as panic attacks. Behavioural and electrophysiological studies have also illustrated traits like faster reaction time [20], accentuated fear memory [21], relatively efficient memory encoding and retrieval [22], as well as voluntary initiation of mental imagery [23] occurring during inhalation, while the converse occurs during exhalation.

As can be seen, a wide range of animal and human studies involving different imaging modalities suggest that respiratory rhythms as well as the lack of them (during voluntary breath-hold) modulate oscillatory patterns throughout the brain. One broad implication is that breathing and breath-hold subserve more than just gas exchange or the lack of it; they organize neuronal assemblies across brain regions to orchestrate complex behaviors affiliated with cognition and emotion.

However, the effect of different phases of the respiratory cycle (like inhalation, exhalation, inhale-hold, and exhale-hold) on cortical dynamics of the human brain is not well understood yet. This study is envisaged as a precursor to understanding whether these different phases have a distinct influence on the functional connectivity between different cortical regions of the human brain at rest. To the best of our knowledge, this aspect has not been explored so far.

## III. METHODOLOGY

### A. Subject Characteristics

A total of 20 subjects (11 males and 9 females), within the age range of 23-60 years (interquartile range: 10.3 years) volunteered to participate in the study. All participants were right-handed and in self-reported good health, with no signs or history of any acute/chronic respiratory pathologies. Written informed consent was obtained from all participants and participation was based on the understanding that observations would be published.

### B. Study Protocol

Participants were instructed to take a shampoo head bath on the day of data recording. They were also directed to avoid hair oil, hair spray, and any metal accessories. Upon arrival at the data collection venue, subjects were briefed about the protocol, and written consent was obtained. The physiological data recording extended over an approximate duration of 20 minutes and during the entire duration, the subject sat upright and relaxed. The protocol involved normal breathing, paced breathing (PB), and breath-hold sessions as shown in Fig. 1.

Paced breathing was performed at a breathing rate of 2 cycles per minute (cpm) for 2 minutes. This was followed by inhale-hold (IH) and exhale-hold (EH) phases, each of 2 minutes duration, at a breathing rate of 2 cpm. During the IH phase, the subject inhales for 7.5 seconds, holds for the following 15 seconds, and then exhales for 7.5 seconds. This cycle is repeated for 2 minutes. Similarly, during the EH phase, the subject inhales for 7.5 seconds, exhales for the following 7.5 seconds, and then holds their breath for 15 seconds. The study protocol is shown in Fig. 1.

A visual metronome has been designed and developed in-house to help subjects synchronize their breathing during the paced breathing and breath-hold sessions. The metronome has two concentric circles. The blue inner circle slowly grows in diameter from the centre, to reach the boundary of the outer circle to depict the inhale phase. Similarly, the inner circle contracts from the boundary of the outer circle to the centre to depict the exhale phase. Participants fix their gaze at the centre and synchronize breathing to the metronome. Also, when the circle is static, participants hold their breath. The metronome is uploaded on our lab website and is available for public access by fellow researchers [24]. The study protocol was approved by the Institute Human Ethics Committee (IHEC) of the Indian Institute of Science (IHEC No: 16/20200821).

### C. Hardware description

EEG data were recorded from all subjects using the eego™ mylab system from ANT Neuro. The system constitutes an eego™amplifier and a 61-channel Waveguard™ EEG cap with Ag/AgCl electrodes embedded in an IFCN extended 10-10 standard electrode layout. A conductive clear gel was used to maintain the scalp-electrode impedance within the acceptable range (< 20 kΩ). The EEG electrode placement used in the study is illustrated in Fig. 2. The scalp electrodes were topographically grouped to define 5 cortical regions, namely frontal (F), central (C), parietal (P), occipital (O), and temporal (T). Data were recorded at a sampling frequency of 1 kHz.

Apart from EEG, the respiratory activity of the subject was recorded using a chest belt, and the cardiac activity was recorded using electrodes placed on the wrists and right leg. Also, galvanic skin response was recorded from the second phalanx of the index finger and ring finger of the non-dominant hand of the subject. Photograph of a subject during the administration of the study protocol is shown in Fig. 3.

### D. Data Pre-processing and Labelling

All pre-processing was carried out using EEGLAB version 13.4 - an interactive MATLAB toolbox for processing continuous and event-related EEG and other electrophysiological data [25]. EEG data collected from each participant during PB, EH, and IH phases were imported into EEGLAB and subjected to line noise removal using the CleanLine EEGLAB plugin [26]. This plugin adaptively estimates and removes line noise from EEG channels using multi-tapering and a Thompson F-statistic. Channels containing excessive artifacts were removed from the data to be interpolated later on in the pre-processing pipeline. EEG epochs of 2.5 seconds duration were generated from all phases. The inhale and exhale segments from paced breathing sessions were respectively used to generate inhale (IN) and exhale (EX) epochs. Similarly, breath-hold segments from IH and EH phases were respectively used to generate inhale-hold (IH), and exhale-hold (EH) epochs. Epoch distribution is summarized in Table I.

TABLE I
EPOCH DISTRIBUTION SUMMARY

| Sl.No. | Condition | Label | # 2.5 s Epochs |
|---|---|---|---|
| 1 | Inhale at 2 cpm | IN | 435 |
| 2 | Inhale-hold | IH | 776 |
| 3 | Exhale at 2 cpm | EX | 168 |
| 4 | Exhale-hold | EH | 768 |
| | Total | | 2147 |

Further, these epochs were visually inspected to discard those containing artifacts due to jaw clenching and head or electrode cable movement. The SIFT EEGLAB plugin [27] was used to perform a piecewise detrending on the data to remove any trends. Subsequently, independent component analysis (ICA) using the runica algorithm was performed to remove artifacts embedded in the data without removing the affected data portions [28], and the automatic algorithm

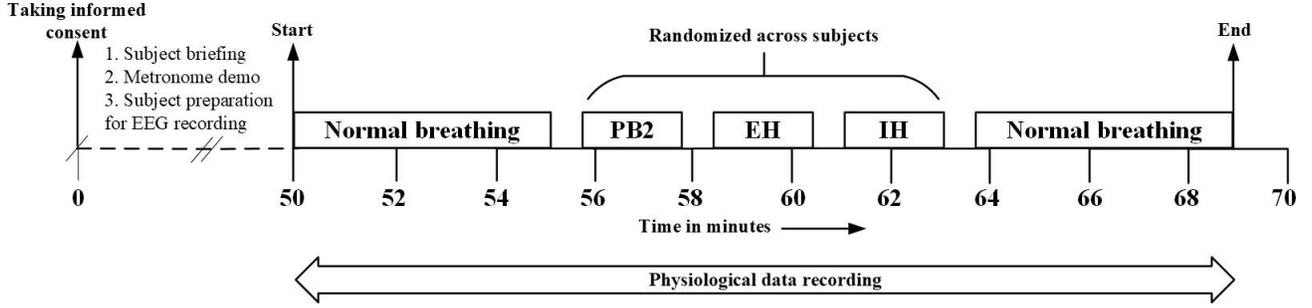

Fig. 1. Data collection protocol. PB2 indicates paced breathing at 2 cpm. EH indicates exhale hold and IH indicates inhale hold.

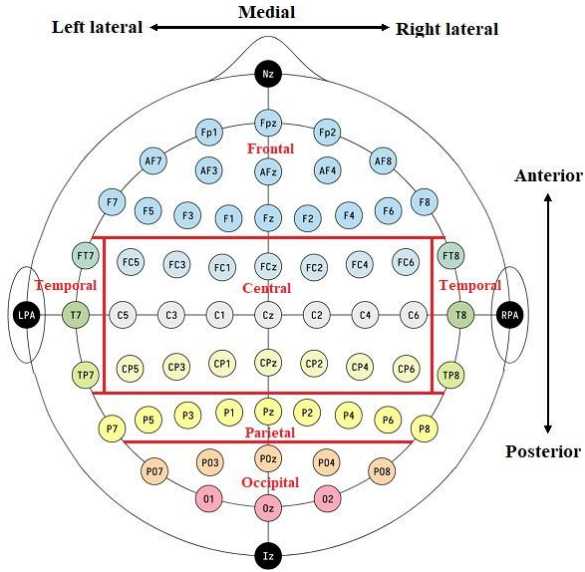

Fig. 2. EEG electrode positions in the 10-10 system. Each electrode placement site has a letter to identify the lobe, or area of the brain: pre-frontal (Fp), anterior frontal (AF), frontal (F), fronto-central (FC), temporal (T), central (C), temporo-parietal (TP), centro-parietal (CP), parietal (P), parieto-occipital (PO), and occipital (O). A "z" refers to an electrode placed on the midline sagittal plane of the skull. Nz denotes the nasion which is the distinctly depressed area between the eyes, just above the bridge of the nose, and Iz is the inion, which is the crest point of back of the skull. LPA and RPA denote the left and right pre-auricular points. Anterior (in front of; towards the face), Posterior (behind; towards the back), Lateral (towards the side), and Medial (towards the middle) are the directional terms.

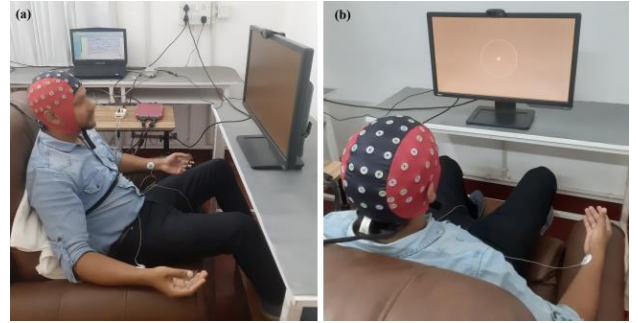

Fig. 3. An illustrative photograph of breath data collection protocol being administered on a male subject. (a) Subject following a visual cue while breathing. Subject may be seen wearing an EEG cap, ECG electrodes, GSR electrodes and chest belt. (b) Visual metronome displayed on the screen.

ADJUST to identify independent components associated with artifacts [29]. The utility of the algorithm in detecting components with artifacts using artifact-specific temporal and spatial features has been validated [29]. After removing artifact-specific independent components selected by the algorithm, data were reconstructed for each phase. Furthermore, the data from previously removed channels were interpolated using a spherical spline interpolation method [30]. The clean EEG data, thus generated, were used for further analyses.

### E. Functional Connectivity Estimation

Functional connectivity (FC) of the brain is operationally defined as the statistical dependence between spatially distant cerebral signals over time [31]. When signals from two anatomically separated brain regions exhibit a synchronization in the time or frequency domain, those regions are said to be functionally connected. This study utilizes the phase slope index (PSI) which is a frequency domain bivariate FC measure that estimates the consistency of the phase lag (or lead) between two signals as a function of frequency [32]. A value that is deviating substantially from zero, for a wider frequency range, suggests that one of the signals is consistently leading the other, thereby indicating an interaction between them.

PSI is mathematically defined as

$$\Psi_{ij} = Im \cdot \sum_{f \in F} C^*_{ij}(f) C_{ij}(f + \delta f) \quad (1)$$

where

$$C_{ij}(f) = \frac{S_{ij}(f)}{\sqrt{S_{ii}(f) S_{jj}(f)}} \quad (2)$$

is the complex coherency, and $\delta f = 1/T$ is the frequency resolution. $Im(.)$ denotes the imaginary part, and $F$ is the frequency range over which the summation is executed.

In this study, $F$ was restricted to 7 specific EEG bands, namely theta (4-8 Hz), alpha1 (8-10 Hz), alpha2 (10-12 Hz), beta1 (12-18 Hz), beta2 (18-21 Hz), beta3 (21-30 Hz), and low gamma (30–45 Hz). PSI was estimated for all possible pairs of EEG time series data recorded from 61 scalp electrode locations.

### F. Feature Selection and Classification

The identification of a brain-scale functional network that can effectively discriminate different phases of the breathing and breath-hold cycle was formulated as a 4-class classification problem with IN, IH, EX, and EH as target classes. The feature space for the classification problem comprised 1830 functional connections from 61 EEG electrode sites.

An optimal and most relevant subset of features was generated using a genetic algorithm (GA). Genetic algorithms are stochastic means for function optimization that aim to replicate natural genetics and biological evolution whereby the genetics of individuals best suited to the environment persist over time [33]. GA is known to present an appealing strategy to identify a near-optimal subset of features by employing bioinspired operators like mutation, crossover, and selection [34].

The classification was performed using a random committee classifier (RCC) [35]. RCC involves a diversified ensemble of base classifiers constructed using the same data but differ in the structure of their models. The randomizable classifier is created using a pseudo-random seed and the final classification result is obtained as the average of the predictions generated by the individual base classifiers.

The implementation involved initializing a random population which is run through the fitness function for the GA. The fitness function returns the best parents. Subsequently, they will be put through the crossover and mutation functions respectively. Crossover is created by combining genes from the two fittest parents by randomly picking a part of the first parent and a part of the second parent. The mutation is achieved by randomly flipping selected bits for the crossover child. A new generation is created by selecting the fittest parents from the previous generation and applying cross-over and mutation. This process is repeated for n number of generations.

For this study, the number of generations was set as 16 and the mutation rate was set to 0.2. About 50% of the fittest individuals would survive into the next generation, and the survival rate of less fit individuals was set to 0.2. The following stopping condition was set to stop the evolutionary process: when the fitness score of a phase is improving only by less than 10%, in comparison to previous phases, the evolving process stops, and the individual with the best fitness is returned. The implemented RCC comprises an ensemble of 10 random trees.

### G. Graph Theoretical Analyses

PSI-based directed and weighted adjacency matrices were formed using functional connections of specific EEG sub-band that could most efficiently discriminate the target classes. These were further used to generate mean brain functional networks uniquely associated with each class. EEG electrodes formed the network nodes and entries of the adjacency matrix formed the edges.

Graph measures like average clustering coefficient (C), average shortest path length (L), and degree assortativity coefficient (A) were used to describe the higher-level architecture of the generated brain functional networks [36]. While C quantifies the degree to which nodes in a graph tend to cluster together, L is the average length of shortest paths for all possible node pairs in the graph, giving an expected distance between two randomly chosen nodes, and A measures the similarity of connections in the graph with respect to the node degree. These measures intuitively characterize how big (or small) the world represented by the network is.

Further, the small-world phenomenon of these networks was evaluated using two small-world coefficients viz., sigma ($\sigma$) and omega ($\omega$) [37], [38]. $\sigma$ is obtained by dividing the normalized C by the normalized L. The normalization is done by dividing the actual value of each measure by the value of the measure for an equivalent random graph. $\omega$ is computed as $(L_r/L - C/C_l)$, where $L_r$ is the average shortest path length of an equivalent random graph and $C_l$ is the average clustering coefficient of an equivalent lattice graph.

### H. Statistical Analyses

The statistical significance of the difference between related groups was estimated using Friedman's ANOVA - a non-parametric version of repeated-measures ANOVA [39]. Wilcoxon post hoc tests with Bonferroni adjustment were used to look for differences between every pair of variables once the main test returned statistical significance.

## IV. RESULTS AND DISCUSSIONS

As discussed in section III-E, whole-brain FC constituting 1830 connections from 61 scalp electrode sites were estimated during IN, IH, EX, and EH phases of the respiratory cycle. Estimations were done separately for 7 specific EEG bands and Fig. 4 summarizes the grand average of these computations over all epochs from all subjects. Friedman ANOVA tests performed on specific EEG bands demonstrated that statistically significant differences existed between target classes in all bands except alpha1. Wilcoxon post hoc tests done on these statistically significant groups pinpointed pairwise groups that showed significant differences. Results are indicated as asterisk brackets in Fig. 4. These results are consistent with our hypothesis that the human brain processes different phases of the respiratory cycle like inhalation, exhalation, inhale-hold, and exhale-hold differently, and the functional connectomes are different during these phases.

Further, these functional differences were explored using feature selection and classification as detailed in section III-F. GA was used to identify a subset of functional connections that would best distinguish the target classes using a RCC, for each EEG band separately. Table II summarizes the results. #FCs indicate the number of functional connections (out of 1830 connections) that yielded the best 10-fold cross-validation (CV) accuracy reported in Table II.

The best accuracy of 93.36 % was obtained when 1161 theta-band functional connections were fed as input to the RCC. Performance measures of this best-performing classification model for each target class are illustrated in Table III. True positive (TP) rate is the proportion of correct positive predictions among the total number of positives, and false positive (FP) rate is the proportion of incorrect positive predictions among the total number of negatives. Precision describes the proportion of correct positive predictions among the total number of positive predictions, and F1 score is the harmonic mean of precision and TP rate.

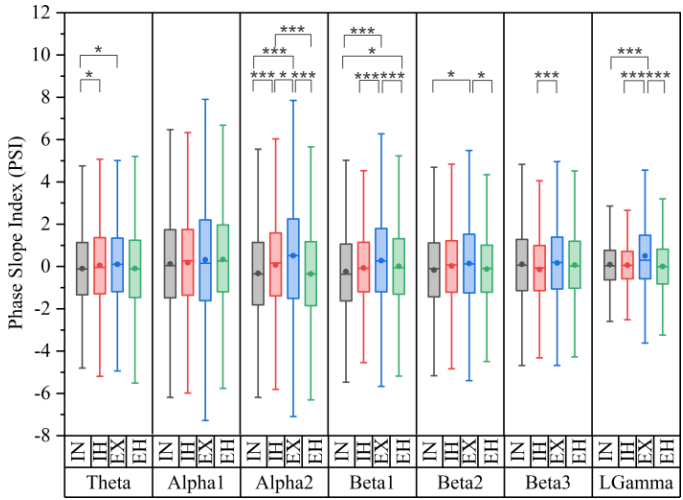

Fig. 4. Grand averages of brain-scale FC estimations during IN, IH, EX, and EH phases for 7 EEG bands are represented as box plots. Box is determined by the 25th and 75th percentiles. Whiskers indicate 1.5 times the interquartile range (IQR). Circle represents the mean value and dash represents the median value. Representation of significance threshold ($p$) is as follows: * is $p \leq 0.05$, ** is $p \leq 0.01$, and *** is $p \leq 0.001$.

TABLE II
CROSS-VALIDATION RESULTS OF RANDOM COMMITTEE CLASSIFIER USING GENETIC ALGORITHM-SELECTED FUNCTIONAL CONNECTIONS AS FEATURES FOR DIFFERENT EEG BANDS

| EEG band | #FCs | CV Accuracy (%) | Kappa |
|---|---|---|---|
| Theta | 1161 | 93.36 | 0.9115 |
| Alpha1 | 988 | 68.17 | 0.5756 |
| Alpha2 | 917 | 66.30 | 0.5506 |
| Beta1 | 900 | 66.24 | 0.5498 |
| Beta2 | 898 | 67.82 | 0.5709 |
| Beta3 | 907 | 67.40 | 0.5653 |
| Low Gamma | 920 | 68.82 | 0.5842 |

TABLE III
PERFORMANCE METRICS OF THE THETA-BAND FC NETWORK THAT BEST DISCRIMINATES THE IN, IH, EX, AND EH PHASES

| EEG band | Theta | | | |
|---|---|---|---|---|
| Cross-validation accuracy | 93.36 % | | | |
| Label | IN | IH | EX | EH |
| TP Rate | 0.910 | 0.936 | 1.000 | 0.888 |
| FP Rate | 0.002 | 0.042 | 0.000 | 0.045 |
| Precision | 0.994 | 0.881 | 1.000 | 0.868 |
| F1 score | 0.950 | 0.907 | 1.000 | 0.878 |

The predominance of the theta band in effectively discriminating phases of breathing and breath-hold may be seen in the context of existing literature that illustrates a statistically significant increase in the theta band spectral power of the frontal, posterior temporal, and occipital regions of the brain during slow breathing in comparison to normal breathing [40]. Theta rhythms possibly originate in the frontal midline regions including the prefrontal cortex and anterior cingulate cortex of the brain [41], [42]. Therefore, our results seem to be in agreement with the studies that attributed a prominent role for the anterior cingulate cortex in respiratory control and suggested that hypercapnia induced by voluntary breath-holding indeed elicits a change in the anterior brain activity [43]. Also, active breathing is known to involve three phases viz., inhalation, post-inhalation, and exhalation, each generated by a distinct excitatory network [44]. Therefore, it may be hypothesized that holding breath after an active inhalation causes inhibition of the post-inspiratory network responsible for the neuronal control of post-inspiratory activity, and the lateral parafacial region (pFL) associated with active exhalation. An exhale hold, on the other hand, causes inhibition of the preBötzinger neuronal network, which is linked to inhalation. Thus, the differentiation between inhale hold and exhale hold may be attributed to variations in the FC patterns induced by these neuronal activities.

The inter and intracortical distribution of 1161 theta-band functional connections are demonstrated in Fig. 5. Based on the topographical grouping illustrated in Fig. 2, 425 of the total 1830 functional connections may be identified as intracortical, ie., both electrodes are on the same cortical region, and 1405 connections are intercortical. As can be seen in Fig. 5, about 60.7% of the intracortical connections and 64.3% of the intercortical connections are retained in the theta-band functional connectome that could best discriminate IN, IH, EX, and EH phases of the respiratory cycle. The maximum number of functional connections of the discriminating connectome is between the frontal and central cortical regions. It may also be noted that all intracortical connections in the occipital region as well as 93.8% of occipital to temporal connections are preserved.

Further, the grand averages of theta band PSI for these intracortical and intercortical functional connections during different phases were computed over all epochs from all subjects, as demonstrated in Fig. 6. Friedman ANOVA tests performed on sub-groups demonstrated statistically significant differences in the intracortical functional connections of the parietal and occipital regions, as well as the FC, CP, CO, CT, FT, and PT intercortical functional connections, during IN, IH, EX, and EH phases. Taken together, it may be concluded that slow deep inhalation and exhalation as well as exhale-hold and inhale-hold have a direct bearing on the cortical activity of the human brain above the brain stem. It may be inferred that the ventilatory changes associated with these phases can lead to alterations in values of pH, and partial pressure of oxygen ($pO_2$) and carbon dioxide ($pCO_2$), thereby causing differential stimulation or inhibition of different cortical regions. The results, therefore, hint at respiration-related entrainment of the cortical activity above the brain stem.

Further, theta band functional networks comprising 1161 edges (GA selected functional connections) and 61 nodes (EEG electrode locations) were generated for the IN, IH, EX, and EH phases. The generated brain functional networks averaged across 20 subjects are illustrated in Fig. 7.

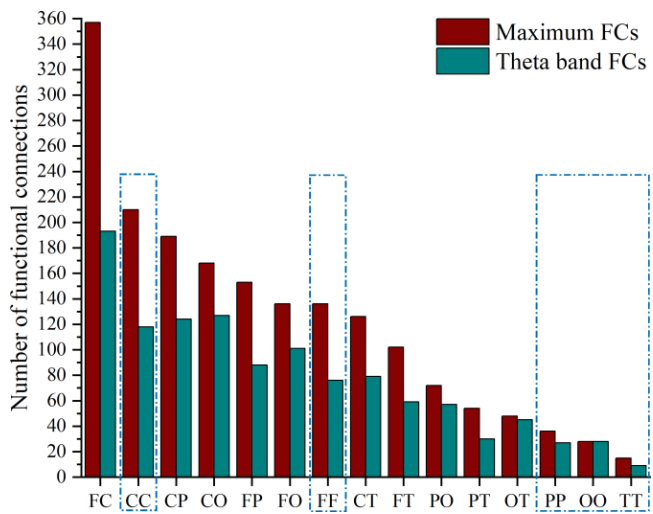

Fig. 5. Intercortical and intracortical distribution of theta band functional connections that best discriminate IN, IH, EX, and EH phases of the respiratory cycle. Inter and intracortical definitions are based on the topographical grouping of EEG electrodes as demonstrated in Fig. 2. Five cortical regions, namely frontal (F), central (C), parietal (P), occipital (O), and temporal (T) are considered. Red bars indicate the total number of functional connections and the green bars indicate the number of selected theta-band connections. Blue dash-dot rectangles highlight the intracortical connections.

These networks were further characterized using graph measures discussed in section III-G. Results are summarized in Table IV.

TABLE IV
GRAPH METRICS OF THE THETA BAND FUNCTIONAL CONNECTOME DURING IN, IH, EX, AND EH PHASES

| Graph Measure | IN | IH | EX | EH | p-value |
|---|---|---|---|---|---|
| Average clustering (C) | 0.071 (0.036) | 0.061 (0.035) | 0.050 (0.025) | 0.049 (0.022) | 1.03E-1 |
| Average shortest path length (L) | 0.117 (0.015) | 0.993 (0.004) | 0.701 (0.239) | 0.405 (0.282) | 4.89E-11 |
| Degree assortativity (A) | -0.102 (0.090) | -0.175 (0.113) | -0.122 (0.088) | -0.195 (0.178) | 8.36E-2 |
| Small-world coefficient ($\sigma$) | 0.997 (0.184) | 0.992 (0.272) | 0.997 (0.206) | 0.968 (0.200) | 7.67E-1 |
| Small-world coefficient ($\omega$) | 0.028 (0.012) | 0.026 (0.010) | 0.025 (0.009) | 0.023 (0.009) | 2.47E-1 |

*Values are mean (SD). p-value gives the significance of the Friedman test.*

Statistical analyses on these metrics highlight that the small-world properties (evaluated by $\sigma$ and $\omega$), the interconnectivity (evaluated by C), and the tendency of nodes to connect to other nodes of similar degree (evaluated by A) were all preserved across the theta-band brain functional networks of different respiratory phases. However, a statistically significant difference in L was observed indicating a difference in the efficiency of information exchange through these networks during IN, IH, EX, and EH phases. Post hoc analyses illustrated significant differences between IN and IH ($Z = 9.699; p < 0.0001$), IN and EX ($Z = 5.889; p < 0.0001$), IH and EX ($Z = 3.811; p < 0.05$), as well as IH and EH ($Z = 6.582; p < 0.0001$). Taken together, our results indicate that theta-band functional connectome of the brain undergoes transitions during different phases of the respiratory cycle and is significantly altered by voluntary breath-holding, and conscious, paced breathing.

## V. CONCLUSIONS

This paper explores the effect of slow, conscious breathing and breath-hold on the functional connectivity of the human brain. The observations highlight that the theta-band brain functional connectome undergoes distinct variations during the inhale, inhale-hold, exhale, and exhale-hold phases of the respiratory cycle. The study also revealed that the network distinction may mainly be attributed to the average shortest path length that distinguishes an easily negotiable network from a complicated and less efficient one. These results may provide interesting clues towards the comprehension of the mechanisms underlying conscious breath control and the extent to which rhythmic brain activity is modulated by the rhythmic act of breathing.

This study, of course, is not without its limitations. It was designed as a proof-of-principle work that aimed to explore the effect of relatively slow, conscious breathing and breath-hold on the functional connectivity networks of the human brain using electroencephalography (EEG). In this study, we only tested a breath rate of 2 cpm so that an inhale-hold and exhale-hold of 15 s, comparable to the inhale and exhale phases of the same duration of slow deep breathing, can be recorded from subjects. Further studies should be performed to optimize the paced breathing and breath-hold durations, as well as other parameters of the protocol, such as the number of intermittent recovery periods. Also, breath-holding time may be increased by training the participants for safe and prolonged breath-holding. Furthermore, the standard and straightforward sensor-space analysis was employed in this study for the functional connectivity estimation. An alternative like the source-space analysis of FC employing methods like cortical partial coherence may be attempted for possibly extracting information about the underlying connections between brain regions.

Since volitional control of breathing by the brain and its effects on cortical dynamics is a closed loop, we believe that our results are a small but substantial step towards understanding these complex dynamics. Since FC is a measure of temporal correlations between brain regions, it can reflect these cortical dynamics, which can further be used to yield graphs of the neural systems uniquely involved in phases of breathing and breath-hold. Identification of such distinct brain graphs may be physiologically and clinically relevant since they help in understanding the cortical pathways involved in breathing and breath-hold, during wakefulness, and conditions associated with oscillatory ventilation such as central apneas (CA) and obstructive sleep apnea (OSA). Furthermore, they can be resourceful in exploring the effects of emerging novel surgical and pharmacological interventions for modulating the chemoreflex control of breathing on cortical activity, in association with various pathological conditions.

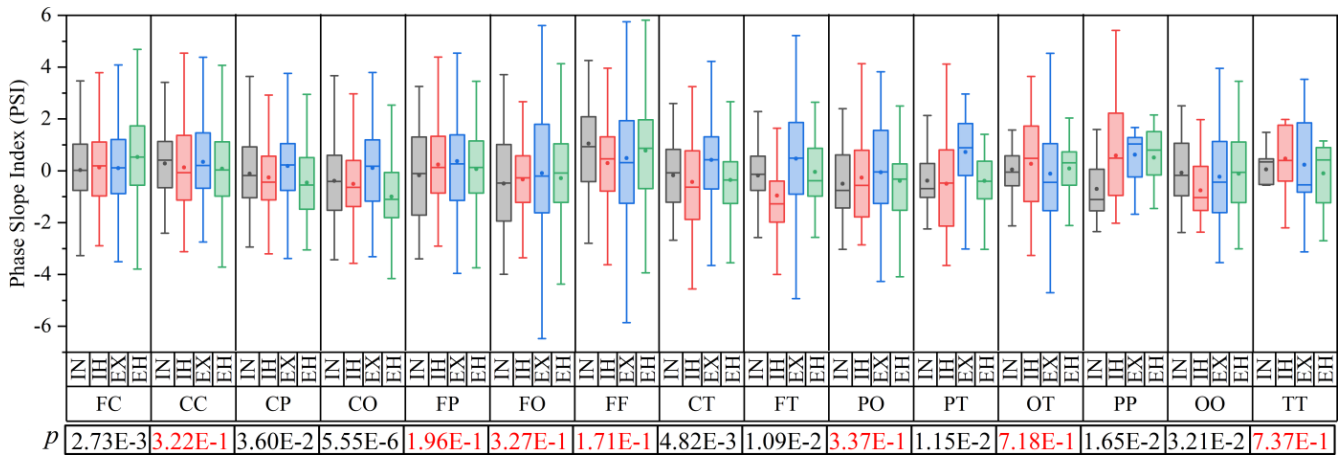

Fig. 6. Grand averages of brain-scale intercortical and intracortical FC estimations for the theta band connectome that best discriminates the IN, IH, EX, and EH phases of the respiratory cycle are represented as box plots. Box is determined by the 25th and 75th percentiles. Whiskers indicate 1.5 times the interquartile range (IQR). Circle represents the mean value and dash represents the median value. The p-values for Friedman's ANOVA tests performed on each subgroup are indicated in the table. Statistically non-significant values are highlighted in red.

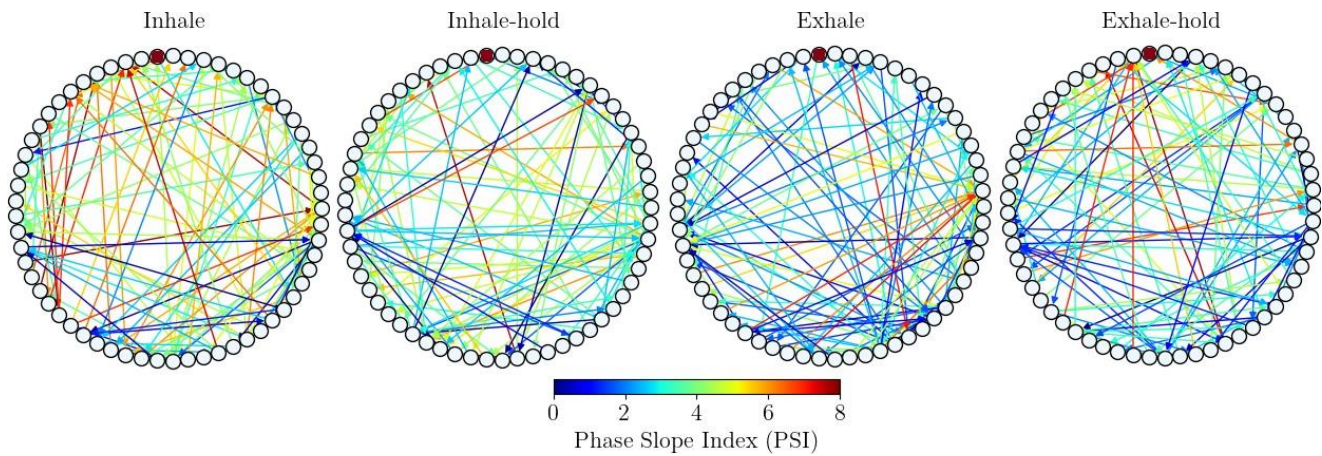

Fig. 7. A representative illustration of theta band brain functional connectome during IN, IH, EX, and EH phases, averaged across 20 subjects. 61 EEG electrodes form the network nodes and PSI entries of the adjacency matrix form the weighted edges. Small circles along the circumference of the circular layout represent nodes of the network. The red filled circle on the top is the Cz node, and clockwise from there are CP6-CP1, C6-C1, AF8, AF7, AF4, AF3, TP8, TP7, T8, T7, Pz, POz, PO8-PO3, P8-P1, Oz, O2, O1, Fz, Fpz, Fp2, Fp1, FT8, FT7, FCz, FC6-FC1, F8-F1 nodes, respectively. Only the strongest incoming and outgoing edges for each node are shown for better visualization.

ACKNOWLEDGMENT

The authors thank the Cognitive Science Research Initiative (CSRI) of the Department of Science & Technology, Government of India for funding part of this work under the postdoctoral fellowship grant No. DST/CSRI-PDF/2021/39 to the first author. The EEG data used for this study was collected under the grant DST/SATYAM/2020/264(G) funded by the Government of India.